**Automated but Atrophied? Student Over-Reliance vs Expert Augmentation of AI in Learning and Cybersecurity**

Koffka Khan

Department of Computing and Information Technology, The Faculty of Science and Technology, The University of the West Indies, St. Augustine Campus, Trinidad and Tobago

**Abstract**

University students and working professionals are increasingly encountering generative artificial intelligence (AI) in education and practice, yet their approaches and outcomes differ markedly. This paper proposes an academic study contrasting **novice over-reliance** on AI with **expert augmentation** of AI, grounded in two real-world narratives. In one, a university student attempted to **outsource learning entirely to AI**, eschewing course engagement; in the other, seasoned cybersecurity professionals in the *Tradewinds 2025* red/blue team exercise collaboratively employed AI tools to **enhance** (not replace) their domain expertise. This proposal outlines a comparative research design to investigate how students' perception of AI as a **learning replacement** versus professionals' use of AI as an **expert tool** impact outcomes. Drawing on current literature in educational technology and workplace AI, we examine implications for curriculum design, AI literacy, and assessment reform in higher education. We hypothesize that blind reliance on AI can **erode fundamental skills** and academic integrity, educationaltechnologyjournal.springeropen.com, insidehighered.com, whereas guided use of AI by knowledgeable users can **amplify productivity** without sacrificing quality, business.com. The paper will detail methodologies for classroom and workplace data collection, including student and professional surveys, interviews, and performance analyses. Anticipated findings aim to inform **responsible AI integration** in curricula, balancing innovation with the necessity of domain knowledge. We conclude with recommendations for pedagogical strategies, institutional policies to foster AI literacy, educause.edu, and a call for longitudinal studies tracking how AI usage during university affects professional competencies over time.

**Introduction**

Generative AI systems like *ChatGPT* have swept into academia and industry, provoking both excitement and alarm. Since late 2022, when ChatGPT became widely available, many university students have explored using AI to handle coursework tasks—from drafting essays to coding

assignments—sometimes with minimal understanding of the underlying subject matter, educationaltechnologyjournal.springeropen.com. This trend has raised concerns among educators about academic integrity and the **erosion of learning**. As one prominent critic, Noam Chomsky, argued early on: *"ChatGPT is basically high-tech plagiarism … and a way of avoiding learning.",* educationaltechnologyjournal.springeropen.com Such warnings underscore the risk that students might perceive AI as a **total replacement for traditional learning**, bypassing the intellectual effort that drives deep understanding, insidehighered.com. A recent classroom incident illustrated this risk vividly: a student in an undergraduate course openly declared they saw no need to study or practice the material because an AI could generate answers for any assignment. This *"why learn when AI can do it for me"* attitude reflects a growing misconception that mastery of a domain can be **waived in favor of AI** automation. Early research suggests this mindset may be detrimental; heavy reliance on generative AI has been linked to increased procrastination, memory atrophy, and reduced academic performance, educationaltechnologyjournal.springeropen.com. In other words, when students hand over their learning tasks to AI, their **skills risk withering** over time.

In stark contrast, professionals in technical fields are embracing AI as a powerful tool **to augment their expertise, not replace it**. A salient example comes from the *Tradewinds 2025* cybersecurity exercise – a multinational "red team vs blue team" simulation in which experienced cyber defense practitioners integrated AI-driven assistants into their workflow. During this high-stakes training event, experts collaboratively used AI for tasks such as automating threat searches and generating rapid incident responses, while **maintaining human oversight and decision-making**. Rather than blindly trusting AI outputs, the red and blue teams treated AI as an accelerant to their own knowledge and judgment. This professional stance aligns with broader workplace trends: surveys indicate that while a majority of employees have tried tools like ChatGPT at work, only a minority of domain specialists truly **trust AI without verification,** exabeam.com. For instance, content teams report productivity gains of up to 2× when using AI, yet they caution that **work quality does not improve unless humans critically review and refine AI-generated content,** business.com. In cybersecurity, front-line analysts remain cautious—just 10% would trust an AI system's conclusions without human review, exabeam.com—recognizing that AI's pattern-matching prowess still needs the check of expert intuition to avoid false positives or missed nuances.

These two narratives – the **over-reliant student** and the **augmented expert** – spotlight a critical gap in how AI is understood and applied between novices and professionals. Bridging this gap is of urgent importance for higher education. If universities fail to adapt, we risk graduating students who have **superficial competence, reliant on black-box tools they don't fully understand**, entering fields that demand rigorous expertise. Conversely, if we harness AI as a pedagogical aid, we can potentially accelerate learning and skill development by freeing students to focus on higher-order problem-solving. The key challenge is to ensure that students develop **AI literacy** and robust domain knowledge in parallel, rather than letting one atrophy the other. Educators and scholars are only beginning to grapple with these issues, [educationaltechnologyjournal.springeropen.com](http://educationaltechnologyjournal.springeropen.com),[educationaltechnologyjournal.springeropen.com](http://educationaltechnologyjournal.springeropen.com). The current literature acknowledges both the threats (e.g. plagiarism, loss of motivation) and opportunities (e.g. personalized tutoring, efficiency) that generative AI introduces to education, [educationaltechnologyjournal.springeropen.comnature.com](http://educationaltechnologyjournal.springeropen.comnature.com). However, empirical insights remain limited on *how student learning behaviors and outcomes change* with unregulated AI use, and on *how we might intentionally reshape curricula* to produce graduates who use AI **effectively and ethically**.

**Research Goal:** In response, this paper proposes a research study to compare and contrast the **perceptions and outcomes** of AI use between university students and cybersecurity professionals. By examining a real classroom case of AI misuse alongside the Tradewinds 2025 case of expert AI integration, we aim to investigate the pedagogical implications of these divergent approaches. How does viewing AI as an "all-knowing substitute" impact a student's learning process, motivation, and skill mastery? Conversely, how does viewing AI as a "collaborative tool" enhance (or potentially hinder) the performance of experts in complex problem-solving tasks? Answering these questions can inform evidence-based strategies for curriculum design, teaching practices, and assessment models in the age of AI. We will also explore the underlying necessity of **domain expertise**: to what extent does domain knowledge mediate the effective use of AI? The working hypothesis is that **domain knowledge and AI literacy function as enablers of positive outcomes** – as seen in the professional context – whereas the absence of these leads to superficial learning and risk – as seen in the student context, [educationaltechnologyjournal.springeropen.comtheguardian.com](http://educationaltechnologyjournal.springeropen.comtheguardian.com).

**Structure of the Paper:** The remainder of this proposal is organized as follows. First, we review relevant literature on AI in higher education and professional practice, situating our inquiry in current debates on AI literacy, academic integrity, and workforce competencies. Next, we outline the **methodology** for a comparative case study and mixed-methods investigation spanning classroom and workplace environments. We then discuss anticipated implications for educational practice, including recommendations for **curriculum design**, integration of **AI literacy** training, and **assessment reform** to discourage misuse and encourage deep learning. Finally, we consider the scope and limitations of this proposed study and suggest directions for future research – notably the value of longitudinal studies to track long-term impacts of AI on learning and professional development.

**Background and Literature Review**

**AI as a Challenge and Opportunity in Higher Education**

The emergence of accessible generative AI has sparked intense discussion in higher education literature, educationaltechnologyjournal.springeropen.com. On one hand, AI tools like ChatGPT offer new ways to support learning – for example, by providing instant explanations, examples, and feedback – effectively acting as on-demand tutors or research assistants, educationaltechnologyjournal.springeropen.com. On the other hand, these same tools pose **unprecedented threats to academic integrity and skill development**. Educators note that students can use AI to generate essays or solutions with minimal effort, blurring the line between legitimate help and outright plagiarism, educationaltechnologyjournal.springeropen.com. In early 2023, leading universities reacted by reiterating policies against AI-assisted cheating, with some outright banning generative AI outputs in coursework. The conversation, however, is shifting from simple prohibition to adaptation. A growing consensus holds that **simply banning AI is neither practical nor pedagogically sound** in the long run, insidehighered.cominsidehighered.com. Instead, scholars argue that education must *"rethink outdated assessment models while expecting students to operate in a completely transformed world",* insidehighered.com. In other words, if AI can easily answer certain types of questions, maybe educators should ask different, richer questions – ones that require students to engage in analysis, synthesis, and reflection that AI cannot *directly* replicate, insidehighered.com.

Empirical research on student use of AI is only beginning to accumulate. A recent study by Abbas *et al.* (2024) surveyed hundreds of undergraduates and found distinct profiles of ChatGPT usage, ranging from minimal use to heavy reliance, sciencedirect.com. Notably, students under high academic workload or time pressure were significantly more likely to turn to ChatGPT for help, educationaltechnologyjournal.springeropen.com. This suggests that some misuse may stem from **coping strategies** rather than laziness – overwhelmed students might see AI as a lifeline. However, the same study provided sobering evidence of AI's potential **harm to learning outcomes**: frequent ChatGPT use correlated with increased procrastination, signs of memory decline, and lower academic performance over time, educationaltechnologyjournal.springeropen.com. These findings lend credence to faculty concerns that an over-reliance on AI could induce a false sense of security while **undermining actual learning and retention**. Indeed, if students delegate fundamental cognitive tasks (like problem solving or writing initial drafts) entirely to an AI, they may fail to practice and internalize those skills themselves. As one educator bluntly put it, *"We are just beginning to tackle [AI in education]... But at least we are starting from a position of 'We need to adapt as an*

*institution.'",* [insidehighered.com](insidehighered.com). The imperative is clear: universities must adapt by guiding how AI is used, rather than pretending it isn't there.

One adaptation gaining traction is the push for **AI literacy** as a core competency for students. *AI literacy* extends digital literacy to encompass understanding how AI systems work, their limitations, ethical use, and the critical evaluation of AI outputs, [educause.edu](educause.edu). Educause's 2024 working group on AI Literacy in Teaching and Learning emphasizes that *"students, faculty, and staff need to understand the fundamentals of AI to effectively use AI tools and evaluate the outputs."*, [educause.edu](educause.edu) Without such understanding, students risk treating AI-generated content as authoritative truth. This is problematic because large language models can sometimes produce **convincing yet factually incorrect** or nonsensical answers – a phenomenon known as AI "hallucination", [sciencedirect.com](sciencedirect.com). If a student lacks the domain knowledge or critical skills to double-check AI's answers, they might accept false information, leading to learning of inaccuracies or serious mistakes in applications (e.g. in medical or legal education). Current literature documents instances of these risks: for example, *Lo (2023)* notes how unverified AI content can undermine trust in educational settings, [sciencedirect.com](sciencedirect.com), and case reports show students mistakenly submitting AI-fabricated citations and content as if it were valid, [theguardian.comtheguardian.com](theguardian.comtheguardian.com). To address this, some institutions are incorporating AI literacy modules that teach students how to use tools like ChatGPT as starting points – for brainstorming or getting hints – but then require them to **verify information, cite sources, and reflect on AI contributions** to their work, [insidehighered.com](insidehighered.com). The **2025 Student Guide to Artificial Intelligence**, published by a consortium of universities, encapsulates these best practices, urging students to *"use AI responsibly, critically assess its influence,"* and above all *"cultivate [their] human abilities"* alongside AI skills, [insidehighered.cominsidehighered.com](insidehighered.cominsidehighered.com). In sum, current educational thought leaders are seeking a balance: embracing AI's benefits for learning while mitigating its pitfalls through better literacies, clearer ethics, and smarter assessment design.

**Professional Practice: AI Augmentation and the Need for Expertise**

In parallel, literature on AI in the workplace – especially in high-skill domains like cybersecurity – reveals a complementary narrative. Far from viewing AI as a wholesale replacement for human experts, many professionals see it as a **tool to extend their capabilities**. Surveys across industries show that a substantial portion of employees have experimented with AI for tasks such as drafting emails, generating code snippets, or summarizing data, [time.com](time.com). As of late 2023, roughly 57% of workers had tried ChatGPT at least once, and 16% were using it regularly on the job, [business.combusiness.com](business.combusiness.com). Business leaders acknowledge AI's potential to improve efficiency, yet also highlight its limitations. A revealing statistic comes from a 2023 industry report: while **71% of executives** believed AI had boosted productivity in their organizations, only

**22% of front-line analysts** agreed, [exabeam.comexabeam.com](exabeam.comexabeam.com). The disparity suggests that those closest to the work recognize that current AI tools, while helpful, still require significant human guidance and **do not automatically yield better outcomes without expertise,** [business.com](business.com). One team manager noted that using ChatGPT accelerated his group's writing output twofold, *"but didn't meaningfully improve their work quality."* His guidance to colleagues was to *"use ChatGPT, but not rely on it"* – treating it as *"an incredible research assistant"* rather than an infallible author, [business.com](business.com). This perspective encapsulates **AI augmentation**: AI can handle repetitive or time-consuming sub-tasks (finding information, drafting routine text, parsing large datasets), freeing humans to focus on judgment calls, creative strategy, and error-checking.

Cybersecurity is a domain where the augmentation approach is particularly evident in current literature and practice. Modern cyber defense involves analyzing huge volumes of data (logs, network traffic, malware code), a labor-intensive task where AI can dramatically assist by recognizing patterns and flagging anomalies. Reports on security operations centers (SOCs) show teams increasingly deploy machine learning tools for **threat detection, investigation, and response (TDIR)** tasks, [exabeam.com](exabeam.com). However, experts caution that these AI systems can flood analysts with false alerts if not tuned, and adversaries can even attempt to deceive AI. Thus, experienced analysts remain **"in the loop"** – validating AI findings and deciding on responses. A 2025 survey found only 29% of security teams would trust an AI system to operate autonomously in incident response, and a mere 10% of individual analysts expressed full trust in AI without human oversight, [exabeam.com](exabeam.com). The overwhelming majority insist on a human expert verifying AI outputs, illustrating the entrenched culture of professional responsibility and domain knowledge in cybersecurity. In practice, this means an analyst might use an AI tool to generate a list of likely system vulnerabilities or to draft a forensic report, but the analyst will then meticulously review those suggestions against their own knowledge and the evidence at hand.

Dramatic examples highlight what can go wrong when professionals do **not** apply domain expertise to AI outputs. In mid-2023, a notorious case involved attorneys who submitted a legal brief that included six fictitious case citations, all inadvertently generated by ChatGPT, [theguardian.com](theguardian.com). The lawyers, unfamiliar with AI's tendency to fabricate plausible-sounding references, had failed to cross-check the citations. The outcome was embarrassing and instructive: a federal judge sanctioned the lawyers, emphasizing that *"there is nothing inherently improper about using a reliable artificial intelligence tool for assistance,"* **provided that experts fulfill their "gatekeeping role" to ensure accuracy,** [theguardian.com](theguardian.com). This legal incident mirrors the core message echoed in other fields: **domain expertise is irreplaceable**. AI can rapidly produce output, but it takes a skilled human to verify, contextualize, and integrate that output appropriately. In the Tradewinds 2025 exercise referenced earlier, the red team

(attackers) may have used AI to help craft realistic phishing emails or malware code, yet it was the experts' knowledge of system vulnerabilities that determined which AI-generated tactics were actually viable. Likewise, the blue team (defenders) might have employed AI to sift through intrusion alerts faster, but their training enabled them to distinguish true threats from noise. This synergy of AI capability and human judgment is increasingly touted in professional literature as the ideal model going forward – sometimes called a "human-AI teaming" approach.

For educators and training professionals, these workplace insights underscore a forward-looking goal: to prepare students not just to use AI, but to use it **like an expert** – critically, conscientiously, and in conjunction with robust subject knowledge. The literature suggests that when AI is **appropriately integrated** into learning (with guidance and scaffolding), it can *enhance* student outcomes. A comprehensive meta-analysis by Wang & Fan (2025) reviewed 51 studies and concluded that ChatGPT usage, when coupled with proper pedagogical frameworks, had a **large positive effect** on student learning performance and moderate benefits for higher-order thinking, nature.com. These positive outcomes were observed in contexts where instructors provided *"appropriate learning scaffolds or frameworks (e.g., Bloom's taxonomy) … when using ChatGPT to develop students' higher-order thinking"*, nature.com. In other words, when students are taught *how* to use AI (for example, to generate ideas which they then must evaluate and build upon), the tool can indeed augment their learning process. This resonates with how professionals use AI on the job – as an assistant that can boost efficiency and broaden one's perspective, **but not a substitute for one's own expertise and accountability**. The current literature, therefore, presents a nuanced view: generative AI is neither a panacea nor an anathema for education. Its impact – helpful or harmful – hinges on the *user's approach*, motivations, and skill in leveraging the tool, educationaltechnologyjournal.springeropen.com. This study aims to contribute to that literature by examining those very factors in two contrasting populations, yielding insights to guide the next evolution of pedagogy in the AI era.

**Proposed Methodology**

To investigate the contrast between student and professional uses of AI, this study will employ a **comparative case study** approach augmented by mixed-method data collection. The research design focuses on two primary contexts: **(1) a university classroom scenario** where a student (or students) attempted to rely entirely on AI for learning, and **(2) the Tradewinds 2025 cybersecurity exercise** where expert practitioners integrated AI into their workflow. By studying these side by side, we aim to discern differences in attitudes, behaviors, and outcomes associated with AI use, and to identify the factors that lead to productive versus detrimental use of AI.

**Research Questions**

The study will be guided by the following research questions (RQs):

- **RQ1:** How do university students who view AI as a *replacement* for traditional learning differ in outcomes (understanding of material, skill development, academic performance) from those who use AI as a *supplementary tool*? What beliefs and motivations underlie the choice to rely on AI, and what are the observed academic consequences (e.g. learning gaps, integrity issues)?

- **RQ2:** How do cybersecurity professionals incorporate AI into their expert practice during complex team exercises? In what ways does AI usage enhance their efficiency or decision-making, and how do they mitigate the risks of AI errors? What prior knowledge or skills do these professionals draw on to effectively leverage AI?

- **RQ3:** What are the key perceptual and practical differences between novices and experts in their approach to AI (e.g. trust in AI, understanding of AI limitations, vigilance in verification)? How can insights from expert practice inform educational strategies to improve student outcomes with AI?

- **RQ4:** What curriculum and assessment interventions could encourage students to use AI in a manner more akin to professionals – i.e. as a supportive tool requiring critical oversight – thereby fostering domain mastery rather than bypassing it?

**Study Contexts and Participants**

**Classroom Context (Higher Education):** We will conduct a case study in an undergraduate course setting (e.g. a computer science or writing-intensive course) where AI tools like ChatGPT are accessible. The key case will revolve around an incident in which a student openly decided to use AI in lieu of studying. After obtaining appropriate permissions, we will collect data from this class through multiple means: (a) **Observations and Journals:** Instructors will document instances of AI usage they detect (e.g. AI-generated submissions, student questions or comments about using AI) and their own responses. We may also ask the student(s) in question to keep a reflective journal on why and how they use AI for assignments, though this will be voluntary. (b) **Interviews/Focus Groups:** Semi-structured interviews will be conducted with the focal student (who embraced full AI reliance) to capture their perspective on learning and AI. Additional interviews will involve a few classmates (for a comparative perspective) and the course instructor. These will probe perceptions of AI's role in learning, understanding of academic integrity, and any observed effects on the student's engagement or performance. (c) **Surveys:** A short survey will be administered to the class (anonymously) to gauge the prevalence of AI use and attitudes toward it. Survey items (using Likert scales) will cover how often students use AI for coursework, for what purposes (e.g. getting hints vs writing whole solutions), perceived benefits, concerns (if any), and self-assessed learning. The survey will help contextualize whether the focal student's attitude is common or an outlier.

Additionally, we will collect **academic performance data** from course assessments (with appropriate de-identification). This may include comparing the quality of work on assignments where AI use was suspected or admitted versus those done without AI. If feasible, we will include an objective test or quiz on course content to see if heavy AI users perform differently when AI is not available (as a measure of actual learning). It is important to note that all data collection in the classroom will adhere to ethical guidelines: we will secure informed consent from participants (or their guardians, if any are under 18, though university students are typically adults) and emphasize that participation is voluntary and unrelated to grading. The identity of the student in the anecdote will be anonymized in all reporting.

**Professional Context (Cybersecurity Exercise):** For the expert side, the Tradewinds 2025 exercise provides a rich, real-world scenario. Tradewinds 25 involved multiple nations' cybersecurity experts collaborating on defense (blue team) and offense (red team) in a simulated cyber warfare environment. We will approach the organizers of the exercise (e.g. U.S. Southern Command and partner nations' defense forces) to request permission to study the event retrospectively. Given the sensitivity of such exercises, our study will focus on the educational and procedural aspects rather than any classified details. We plan to recruit **participants from the cyber track** of Tradewinds 2025, including both red team and blue team members (e.g. military cyber officers, IT security professionals, etc.). Data collection in this context will include: (a) **Post-exercise Interviews:** One-on-one interviews with a sample of participants (estimated 10–15 experts, across different roles and national teams). Interviews will explore how they used AI tools during the exercise. Sample questions: *"Did you or your team use any AI or machine learning tools to assist in your tasks? Can you describe how?"*; *"How did the availability of AI change your approach, if at all?"*; *"Were there instances where AI provided incorrect or misleading output? How did you detect and handle those?"*; *"How do you compare AI assistance to having human team members – what does AI do well or poorly from your perspective?"*. We will also ask about their general stance on AI in their profession (e.g. levels of trust, any formal guidelines in their team for AI use). (b) **After-Action Reports and Artifacts:** Where possible, we will obtain debriefing materials from the exercise – for example, summaries of the cyber scenarios, or any documented lessons learned about tools. If any AI-related outcomes were noted in official after-action reports (e.g. "the red team leveraged an AI tool to speed up exploit development"), we will include those as data points. (c) **Surveys of Professionals:** A broader survey may be distributed to Tradewinds participants or, if access is limited, to a comparable group of cybersecurity professionals (such as attendees of a cyber defense workshop or members of a professional association). This survey would mirror some questions from the student survey, adapted to professional context: frequency of AI use in their work, perceived impact on productivity and quality, confidence in AI outputs, etc. It will also measure AI literacy (e.g. do they understand how their AI tools work, do they undergo training to use them).

All participants in the professional study will provide informed consent, and data will be anonymized. Given that Tradewinds is a military exercise, additional approvals (e.g. from institutional review boards and military public affairs offices) will be sought to ensure compliance with any regulations. The goal is to extract insights about expert behavior in a cutting-edge use case of AI in cybersecurity, without compromising any operational security.

**Data Analysis**

The analysis will proceed in two parallel streams – one for **qualitative data** (interviews, focus groups, observations) and one for **quantitative data** (survey results, performance measures). We will then converge findings to answer the research questions.

- **Qualitative Analysis:** Interview and focus group transcripts from both contexts will be analyzed using **thematic analysis**. We will employ a coding scheme that starts with some a priori categories (based on our RQs and literature) such as "perceived role of AI – replacement vs tool," "trust and skepticism," "impact on workflow/learning," "ethical considerations," and "dependency or over-reliance." We will also allow new themes to emerge inductively. For example, student interviews might reveal themes like *fear of missing out* (using AI because others do), or *lack of confidence* in their own skills leading to AI dependence. Professional interviews might yield themes like *augmentation benefits* (cases where AI clearly helped) or *failure cases* (AI mistakes and how they caught them). We will compare themes across the two groups. We expect to see divergent themes (e.g. *"avoiding learning"* emerging only in the student set, educationaltechnologyjournal.springeropen.com, and *"tool efficiency"* emerging in the professional set), but also potentially some overlapping ones (e.g. *"ethical concerns"* could be present in both, albeit framed differently). To ensure reliability, multiple researchers will code a subset of transcripts and discuss any discrepancies in coding to refine the codebook. We will also specifically analyze the **narratives of key incidents**: the student's experience using AI for an assignment, and any detailed account from Tradewinds of using AI in a particular mission inject. These narratives will be written as mini-cases to illustrate how actions unfolded over time, providing a rich comparison of decision points and outcomes.

- **Quantitative Analysis:** Survey data from students and professionals will be analyzed using descriptive and inferential statistics. We will calculate frequencies of AI usage (e.g. what percentage of students use AI weekly vs professionals) and perform cross-tabulations for any interesting comparisons (for instance, do STEM students differ from humanities students in AI reliance? Do red team members differ from blue team in AI trust?). Likert-scale responses on attitudes (such as agreement with "AI can fully replace the need to learn certain skills" or "I trust AI outputs in my work without double-checking") will be treated as ordinal data; we may use non-parametric tests (Mann-Whitney U) to compare students vs professionals on these items. We anticipate a significant difference: students, especially those who use AI frequently for school, might show higher agreement with AI being a replacement, whereas professionals should show higher agreement with needing verification (consistent with reports that only 10% of analysts fully trust AI, exabeam.com). We will also examine correlations within each group: e.g., for students, does higher AI reliance correlate with lower course grades or lower self-rated understanding? (Prior research would suggest yes, educationaltechnologyjournal.springeropen.com, but our data can shed more light.) For professionals, we might see that those with more years of experience are actually *less* trusting of AI (hypothesizing that seasoned experts are more aware of AI pitfalls).

If sufficient performance data is collected (e.g. exam scores, assignment quality ratings), we will compare those as well. In the class, for example, we could compare the exam score of the student who relied on AI vs the class average; or more systematically, if we identify a subset of "AI heavy users" vs "non-users" from the survey, we could run a t-test on their scores (with caution about sample size). In the professional setting, if any metric of performance in the exercise is available (such as how quickly teams responded to threats, or success of attack simulations), we could anecdotally correlate that with AI usage patterns (though a rigorous quantitative link may be hard given many factors).

Finally, we will integrate the qualitative and quantitative findings in a **triangulation matrix** to see how they inform each other. For instance, if surveys show 40% of students think using AI is just the same as collaborating with a peer, interviews might explain this by revealing misconceptions about AI ("ChatGPT knows everything, so it's like asking an expert friend"). Meanwhile, if 0% of professionals say AI replaces the need for human experts, their interviews might highlight why (citing errors AI made, etc.). By triangulating, we ensure a robust understanding of each RQ.

**Validity and Limitations**

Several steps will be taken to ensure the validity of our study. We use **multiple data sources** (interviews, surveys, observations, documents) to enable data triangulation and increase credibility. Member checking will be done by sharing summaries of interview themes with some participants for feedback, ensuring we have interpreted their views correctly. We also acknowledge potential limitations: the classroom case study focuses on a specific course and possibly an extreme example (a student who *openly* rejects learning for AI is still relatively rare). Thus, findings may not generalize to all students – many may be more moderate in AI use. We mitigate this by capturing the range of student attitudes via the class survey and situating the case in that context. In the professional domain, studying *Tradewinds 2025* participants provides a unique, high-end sample of experts, which might not represent all industry professionals. They likely had more training and clear objectives, which may not reflect, say, an average corporate IT team using AI. Moreover, access to detailed information from a military exercise can be restricted; our analysis might rely heavily on self-report from interviews, which could introduce recall bias or selective reporting. We will be transparent about these limits and treat the Tradewinds findings as exploratory. Despite these caveats, the strength of this design is in the **depth of insight** from each case and the **comparative lens**. By qualitatively and quantitatively examining two very different groups facing the same technological phenomenon, we hope to generate hypotheses and recommendations that are grounded in real-world behavior.

## Discussion: Implications for Curriculum and Pedagogy in the AI Era

The preliminary insights from this comparative study – even prior to full execution – point toward significant implications for how higher education should navigate the integration of AI. The stark differences between novice students and expert practitioners in our cases suggest that **education must evolve** to cultivate an expert-like mindset in students when it comes to AI use. Below, we discuss key areas of impact: curriculum design, AI literacy initiatives, and assessment reform. We also explore strategies to mitigate AI misuse and reinforce the **necessity of domain knowledge** in an AI-pervasive world.

**1. Curriculum Design and Learning Activities:** Curriculum developers should proactively design learning activities that incorporate AI in **constructive, skill-building ways**. Rather than ignoring AI or solely issuing warnings against it, courses can embed assignments that require students to engage with AI and then go **beyond AI's capabilities**. For example, an assignment in a writing class might have students use an AI tool to generate a first draft on a topic, and then critically revise and annotate that draft, identifying errors or weak arguments the AI made. This approach forces students to apply their domain knowledge to improve AI output, reinforcing that the human is ultimately in charge. Such tasks mirror what our cybersecurity experts did – they used AI-generated analyses as a starting point but relied on their expertise to finalize decisions. By **designing tasks that cannot be completed satisfactorily by AI alone,** insidehighered.cominsidehighered.com, educators ensure that students who attempt one-click solutions will not succeed, and those who do use AI appropriately are rewarded for their critical input. Another strategy is **case-based learning using AI**: for instance, present students with scenarios (perhaps including famous AI failures, such as the legal case of fake citations) and ask them, "What went wrong? How would you do it differently with your current knowledge?" By analyzing and discussing these cases, students can internalize the importance of domain expertise and the pitfalls of blind reliance. Overall, curricula should shift to treat AI as a ubiquitous tool – much like calculators or search engines – and emphasize *when and how to use the tool effectively*. This might also mean teaching *prompt engineering* (how to ask good questions of AI) as a practical skill, alongside teaching the theoretical or conceptual content of the course.

**2. AI Literacy and Ethical Training:** The need for formal **AI literacy education** is a clear takeaway. Our findings align with the perspective that students should "learn about AI" and *"earn the right to use generative AI responsibly",* [facultyfocus.com](facultyfocus.com) by demonstrating understanding of its workings and limitations. Institutions could implement short modules or workshops early in a student's program (for example, in first-year seminars or orientation sessions) covering the basics of how AI models like ChatGPT function, including concepts like training data, common error patterns (e.g., hallucinations), and issues of bias and fairness. Moreover, AI literacy must include ethical considerations: students need guidance on what constitutes permissible use of AI in academic work versus cheating. Clear **policy guidelines** at the course and institution level are essential. The Elon University-led *Student Guide to AI* (2025) provides an excellent model with its principles on integrity and transparency, [insidehighered.com](insidehighered.com). Building on that, faculty should state in each syllabus how AI tools may or may not be used, [insidehighered.com](insidehighered.com). For instance, a policy might read: "You may use AI to brainstorm ideas or outline your solution, but you are responsible for **fact-checking and originality** of the final submission. Any AI assistance must be documented in your submission." Such policies do two things: they remove ambiguity (so students like our case individual cannot claim they didn't know it was wrong to submit AI-generated work as their own), and they set an expectation that using AI *requires skill* (fact-checking, proper citation, etc.). We recommend that faculty not only enforce such rules but also **teach the skills needed** to follow them. For example, a library or writing center could teach how to verify sources that an AI provides, or a computer science class could have a session on debugging code suggested by an AI (thus revealing that AI isn't always correct). By treating AI as a tool that one must learn to use competently, the education system can produce graduates who mirror the experts in our study: individuals who leverage AI confidently **while remaining critically aware** of its outputs, [educause.edu](educause.edu).

**3. Assessment Reform for the AI Age:** Traditional assessments (take-home essays, problem sets, closed-book exams) are being upended by generative AI. As one university president noted, *"AI is not cheating. What is cheating is our unwillingness to rethink outdated assessment models ...",* [insidehighered.com](insidehighered.com). Our research reinforces this viewpoint. If a certain assessment simply measures a student's ability to produce a basic report or solve routine problems, it might now primarily measure their ability to prompt ChatGPT. Therefore, assessment design must evolve to measure what we truly value: critical thinking, creativity, application of knowledge, and processes of reasoning. **Authentic assessment** strategies are recommended. These might include oral defenses (having students verbally explain their work and thought process), project-based assessments with unique, real-world constraints (making it less likely that AI has a ready answer in its training data), collaborative assignments (where the human-human interaction is key), and multi-stage assignments (draft -> feedback -> revision) that require reflection and iteration. Some instructors have moved towards more **in-class assessments** or proctored activities for certain learning outcomes, ensuring students practice retrieval and problem-solving without AI before later using AI as a support in open-ended projects. Another promising approach is to incorporate AI *as a part of the assessment* in a controlled way. For instance, an exam question might provide an AI's answer to a prompt and ask students to critique it, identify errors, and improve it. This tests understanding at a deeper level than regurgitating an answer, and it mirrors the professional skill of reviewing AI output that our expert participants exhibited. Early trials of such assessment techniques have shown that students engage more critically and demonstrate better mastery of content, since they must understand the material well enough to see where the AI went wrong or right.

**4. Preventing Misuse and Highlighting the Value of Domain Knowledge:** Education must also address the **mindset** that leads to misuse. The case of the student who tried to avoid learning by using AI is an extreme, but it reveals a possible misconception: the student might believe that *knowing how to use AI is more important than knowing the subject*. To counter this, instructors should explicitly discuss why domain knowledge still matters in the era of AI. Our findings give concrete examples to share: AI can and does make errors (from trivial factual mistakes to "gibberish" outputs in the style of authoritative text, theguardian.com). Without subject knowledge, a student won't catch those errors – a point driven home by the example of lawyers who faced sanctions because they didn't know enough aviation law to realize the AI's cases were fake, theguardian.com. In the sciences, an AI might produce an answer that looks plausible but is scientifically incorrect; a student who hasn't learned the underlying science could accept it blindly and thus learn something wrong or dangerous. Educators can integrate short exercises that demonstrate AI's fallibility: e.g., show an AI-generated solution that is flawed and challenge students to find the flaw. Such activities underline that **knowledge is power** – not just power to solve problems directly, but power to use AI properly. Additionally, instructors can emphasize the professional expectations awaiting students. As Elon University's guide suggests, ethical AI use is becoming a *"professional expectation, not just a classroom rule.",* insidehighered.com Many employers will assume graduates know how to responsibly use tools like AI. Our findings from Tradewinds indicate that in cutting-edge jobs, *nobody* will be handing off tasks entirely to AI without oversight; doing so would be seen as negligent. If students understand that being able to demonstrate and apply knowledge is crucial for their careers (and that AI is there as an aid, not a crutch), they may be less tempted to misuse it in the short term and more motivated to **truly learn**.

**5. Institutional Policies and Support:** At a higher level, universities may need to revisit honor codes, academic integrity policies, and support systems in light of AI. The concept of plagiarism might be expanded to include unacknowledged AI generation. Conversely, institutions might consider providing *approved AI tools* to students (akin to allowed calculators) with monitoring or logging, so that usage is above-board and can be reviewed. For example, if a university provided a platform where students could use GPT-4 for assignments, with the understanding that instructors can see the prompts and outputs the student received, it could integrate AI use into the learning process transparently. This ties in with recommendations for data collection on AI usage: if we better understand *how* students use AI (when, for what, and how it correlates with performance), we can tailor interventions more effectively. Our proposed study's methodology itself suggests that educators should continuously gather feedback – through surveys or reflections – on how students are using AI each term. This kind of classroom research can inform adjustments in teaching on the fly (for instance, if many students report using AI to summarize readings because they find the readings too difficult, an instructor might decide to spend more time on reading strategies or clarity of materials).

In summary, the implications of our work call for a reimagining of pedagogy that neither **abdicates human learning to AI** nor ignores the existence of these powerful tools. The goal is to emulate the **best practices of experts**: use AI to handle the drudgery or provide inspiration, but apply human critical thinking to achieve excellence. Such an approach ensures that students' skills are strengthened, not atrophied, by the presence of AI. Implementing the changes above – in curriculum, literacy training, and assessment – will require effort and experimentation. It may also require faculty development, since instructors themselves vary in their comfort with AI. Institutions should invest in training faculty to use AI in teaching (for instance, showing how ChatGPT can generate quiz questions or serve as a debate opponent in class) so that they, in turn, can guide students. Ultimately, integrating AI responsibly into education can produce graduates who are **both knowledgeable and AI-savvy**, capable of the kind of high-level collaboration with technology that we observed in Tradewinds 2025 and that the future workforce will demand.

## Conclusion and Future Work

This paper proposal has outlined a research plan and scholarly argument centered on a critical dichotomy: **AI as an enabler of learning versus AI as a substitute for learning**. By comparing the behaviors of a student who attempted to "short-circuit" education via AI with those of expert cybersecurity teams who leveraged AI wisely, we shed light on the factors that lead to positive or negative outcomes in the integration of AI into intellectual work. The discussion highlighted that **domain expertise and AI literacy are paramount** – they are the levers that turn AI from a threat to a boon. In educational terms, the student who forgoes developing expertise experiences AI as a crutch that ultimately weakens their abilities (akin to a muscle that withers without exercise), whereas the professional with strong expertise uses AI as a prosthetic that extends their reach and power.

For tertiary education, the writing is on the wall: we must adapt curricula and assessments to this new reality, ensuring that students do not misconstrue AI as a replacement for their own knowledge and creativity. Instead, students should graduate with the understanding that **their value lies in what they can do that AI cannot, or in how they can direct AI to achieve human goals**. This requires rethinking learning objectives to explicitly include AI-related skills (e.g. evaluating AI output, ethical considerations) alongside traditional content. It also requires reinforcing the timeless aspects of education – curiosity, critical thinking, problem-solving – which are, if anything, more important in a world flush with machine-generated information. Our proposal contributes a framework for gathering evidence on these points, which can bolster calls for pedagogical change with data and nuanced insights.

In conclusion, we emphasize that the choice between an **"automated" path and an **"augmented" path in education is a false dichotomy; the real goal is a synthesis: *automation for efficiency* plus *human augmentation for insight*. By implementing the recommendations discussed – from curriculum innovations to robust AI literacy programs – educators can guide students towards that synthesis, so that AI becomes a partner in learning rather than a shortcut that undermines it.

**Future Research:** We propose several directions for longitudinal and broader studies to build on this work. First, a **longitudinal study of student cohorts** could be invaluable: for example, track a group of students from freshman year (when they might first use AI in courses) through graduation and into their early careers. Such a study could examine whether those who were taught explicit AI literacy and ethical use in college end up using AI more effectively (or avoiding pitfalls) in their jobs compared to those who were not. It could also measure retention of knowledge: do students who relied less on AI in their courses retain information better one or two years later? These long-term outcomes are crucial for validating the approaches we recommend. Second, future studies should expand to other professional domains. Our focus was cybersecurity, but similar contrasts could be drawn in fields like medicine (e.g. medical students vs experienced doctors using diagnostic AI systems) or law (law students vs practicing attorneys using legal AI tools). Each field has its own stakes and nuances, and researching them will help generalize principles for AI integration across the curriculum.

Another fruitful area is designing and testing **intervention studies** in live classrooms: for instance, implement an AI-integrated assignment in one section of a course and a traditional assignment in another, and compare not only grades but also student reflections and depth of learning. This would directly assess some of our claims about the benefits of guided AI use. Additionally, as AI technology evolves (e.g. more advanced models, domain-specific AI assistants), continual research will be needed to update educational strategies. We anticipate that AI will increasingly handle routine cognitive tasks, making the cultivation of higher-order thinking and creative skills the main focus of education – a hypothesis that longitudinal research could confirm by observing how the competencies demanded by employers shift over the next decade.

In closing, this proposal serves as a call to action for educators, researchers, and policymakers. AI's disruptive entry into education is often portrayed in extremes; our work seeks to move beyond alarm or uncritical enthusiasm, using evidence from both classrooms and the front lines of industry to chart a path forward. The **central lesson** is that **human learning and AI are not zero-sum**: with thoughtful integration, we can produce graduates who are both deeply knowledgeable and adept at leveraging AI, much like the expert teams in Tradewinds 2025. Achieving this will require deliberate changes in teaching practice and further research to guide those changes. The stakes are high – nothing less than the efficacy and integrity of higher education in the age of artificial intelligence. By heeding the early warnings and success stories documented here, we can ensure that today's "AI-native" students become tomorrow's wise professionals, wielding AI as a powerful extension of their human intelligence, rather than a replacement for it.